\documentclass[preprint,aps,showpacs]{revtex4}
\usepackage{mathrsfs}
\usepackage{amsmath}
\usepackage{amssymb}
\usepackage{epsfig}
\usepackage{graphicx}
\usepackage{booktabs}
\usepackage{multirow}

\begin{document}
\renewcommand{\arraystretch}{0.5}
\newcommand{\beq}{\begin{eqnarray}}
\newcommand{\eeq}{\end{eqnarray}}
\newcommand{\non}{\nonumber\\ }

\newcommand{\acp}{ {\cal A}_{CP} }
\newcommand{\psl}{ p \hspace{-1.8truemm}/ }
\newcommand{\nsl}{ n \hspace{-2.2truemm}/ }
\newcommand{\vsl}{ v \hspace{-2.2truemm}/ }
\newcommand{\epsl}{\epsilon \hspace{-1.8truemm}/\,  }

\def \cpl{ Chin. Phys. Lett.  }
\def \ctp{ Commun. Theor. Phys.  }
\def \epjc{ Eur. Phys. J. C }
\def \jpg{  J. Phys. G }
\def \npb{  Nucl. Phys. B }
\def \plb{  Phys. Lett. B }
\def \prd{  Phys. Rev. D }
\def \prl{  Phys. Rev. Lett.  }
\def \zpc{  Z. Phys. C }
\def \jhep{ J. High Energy Phys.  }

\title{ The double charm decays of $B_c$ Meson in the
Perturbative QCD Approach}
\author{Zhou Rui $^{1,2}$}
\author{Zou Zhitian $^1$}
\author{Cai-Dian L\"{u}$^1$}\email{lucd@ihep.ac.cn}

 \affiliation{$^1$ Institute
of  High  Energy  Physics  and  Theoretical  Physics Center for
Science Facilities, Chinese Academy of Sciences, Beijing 100049,
People's Republic of China }

\affiliation{$^2$ School of Science,
 Hebei United University,
Tangshan, Hebei 063009, People's Republic of China }
\date{\today}
\begin{abstract}
We study the double charm decays of $B_c$ meson, by employing the
perturbative QCD approach based on $k_T$ factorization. In this
approach, we include    the non-factorizable emission diagrams and W
annihilation diagrams, which are   neglected  in the previous naive
factorization approach. The former are important in the
color-suppressed modes; while the latter are important in most $B_c$
decay channels due to the large Cabibbo-Kobayashi-Maskawa matrix
elements. We make comparison with those previous naive factorization
results for the branching ratios  and also give out the theoretical
errors that previously missed. We   predict  the transverse
polarization fractions of $B_c\rightarrow
D^{*+}_{(s)}\bar{D}^{*0},D^{*+}_{(s)}D^{*0}$ decays for the first
time.
 A large transverse polarization contribution
that can reach $50\%\sim 60\%$   is predicted in some of the $B_c$
meson decays.
\end{abstract}



\maketitle

\section{Introduction}

Since the $B_c$ meson is the lowest bound state of two different
heavy quarks with open flavor, it is stable against strong and
electromagnetic annihilation processes. The $B_c$ meson therefore
decays weakly. Furthermore, the $B_c$ meson has a sufficiently large
mass, thus each of the two heavy quarks can decay individually. It
has rich decay channels, and provides a very good place to study
nonleptonic weak decays of heavy mesons, to test the standard model
and to search for any new physics signals \cite{iiba}.  The current
running LHC collider will produce much more $B_c$ mesons than ever
before to make this study a bright future.

Within the standard model (SM), for the double charm decays of
$B_{u,d,s}$  mesons, there are penguin operator contributions as
well as tree operator contributions. Thus the direct CP asymmetry
may be present. However,  the double charm decays of $B_{c}$ meson
are pure tree decay modes, which are particularly well suited to
extract the  Cabibbo-Kobayashi-Maskawa (CKM) angles due to the
absented interference from penguin contributions. As was pointed out
in ref. \cite{plb286160} and further elaborated in ref.
\cite{prd62057503,jpg301445,plb555189,prd65034016}, the decays
$B_c\rightarrow D_s^+D^0, D_s^+ \bar{D}^0$ are the gold-plated modes
for the extraction of CKM angle $\gamma$ though amplitude relations
because their decay widths are expected to be at the same order of
magnitude. But this needs to be examined by faithful calculations.

Although many investigations on the decays of $B_c$ to double-charm
states have been carried out
\cite{jpg301445,plb555189,prd73054024,prd564133,pan67,prd61034012,prd62014019,prd493399}
in the literature, there are uncontrolled large theoretical errors
with quite different numerical results. In fact, all of these old
calculations are based on naive   factorization hypothesis, with
various form factor inputs. Most of them even did not give any
theoretical error estimates because of the non-reliability of these
models. Recently, the theory of non-leptonic B decays has been
improved quite significantly. Factorization has been proved in many
of these decays, thus allow us to give reliable calculations of the
hadronic B decays. It is also shown that the non-factorizable
contributions and annihilation type contributions, which are
neglected in the naive factorization approach, are very important in
these decays \cite{cheng}.

 The perturbative QCD
approach (pQCD) \cite{prl744388} is one of the recently developed
theoretical tools based on QCD to deal with the non-leptonic B
decays. Utilizing the $k_T$ factorization instead of collinear
factorization, this approach is free of end-point singularity. Thus
the Feynman diagrams including factorizable, non-factorizable and
annihilation type, are all calculable. Phenomenologically, the pQCD
approach successfully predict the charmless two-body B decays
\cite{plb5046,prd63074009}. For the decays with a single heavy $D$
meson in the final states (the momentum of the $D$ meson is
$\frac{1}{2}m_B(1-r^2)$, with $r=m_D/m_B$), it is also proved
factorization in the soft-collinear effective theory \cite{scet1}.
Phenomenologically the pQCD approach is also demonstrated to be
applicable in the leading order of the $m_D/m_B$ expansion
\cite{09101424,0512347} for this kind of decays.  For the double
charm decays of $B_c$ meson, the momentum of the final state $D$
meson is $\frac{1}{2}m_{B_c}(1-2r^2)$,   which is only slightly
smaller than  that of the decays with a single D meson final state.
The prove of factorization here is thus trivial. The pQCD approach
is applicable to this kind of decays. In fact, the double charm
decays of $B_{u,d,s}$ meson have been studied in the pQCD approach
successfully \cite{dd1,dd2}, with best agreement with experiments.
In this paper, we will extend our study to     these $B_c$ decays in
the  pQCD approach, in order to give predictions on branching ratios
and polarization fractions for the experiments to test. Since this
study is based on QCD and perturbative expansion, the theoretical
error will be controllable than any of the model calculations.

Our paper is organized as follows: We review the pQCD factorization
approach and then   perform the perturbative calculations for these
considered decay channels   in Sec.\ref{sec:f-work}.  The numerical
results and discussions on the observables are given in
Sec.\ref{sec:result}. The final section is devoted to our
conclusions. Some details of related functions and the decay amplitudes
are given in the Appendix.

\section{  Framework}\label{sec:f-work}

For the double charm decays of  $B_c$, only the tree operators of the standard effective weak
Hamiltonian contribute.  We can divide them  into two groups: CKM favored decays with both
emission and annihilation contributions and pure emission type decays, which are CKM suppressed.
For the former modes, the Hamiltonian is given by:
\begin{eqnarray}\label{eq:H1}
\mathcal {H}_{eff}&=&\frac{G_F}{\sqrt{2}}V_{cb}^*V_{uq}[C_1(\mu)O_1(\mu)+C_2(\mu)O_2(\mu)],\nonumber\\
O_1&=&\bar{b}_{\alpha}\gamma^{\mu}(1-\gamma_5)c_{\beta} \otimes \bar{u}_{\beta}\gamma_{\mu}(1-\gamma_5)q_{\alpha} ,\nonumber\\
O_2&=&\bar{b}_{\alpha}\gamma^{\mu}(1-\gamma_5)c_{\alpha} \otimes \bar{u}_{\beta}\gamma_{\mu}(1-\gamma_5)q_{\beta},
\end{eqnarray}
while   the effective Hamiltonian  of the latter modes reads
\begin{eqnarray}\label{eq:H2}
\mathcal {H}_{eff}&=&\frac{G_F}{\sqrt{2}}V_{ub}^*V_{cq}[C_1(\mu)O'_1(\mu)+C_2(\mu)O'_2(\mu)],\nonumber\\
O'_1&=&\bar{b}_{\alpha}\gamma^{\mu}(1-\gamma_5)u_{\beta} \otimes \bar{c}_{\beta}\gamma_{\mu}(1-\gamma_5)q_{\alpha} ,\nonumber\\
O'_2&=&\bar{b}_{\alpha}\gamma^{\mu}(1-\gamma_5)u_{\alpha} \otimes \bar{c}_{\beta}\gamma_{\mu}(1-\gamma_5)q_{\beta},
\end{eqnarray}
where $V (q=d,s)$ are  the corresponding CKM matrix
elements. $\alpha$, $\beta$ are the color indices.  $C_{1,2}$ are Wilson coefficients at renormalization scale $\mu$.
$O_{1,2}$ and $O'_{1,2}$ are the effective  four-quark operators.

The factorization theorem allows us to factorize the decay amplitude
into the convolution of the hard subamplitude, the Wilson
coefficient  and  the meson wave functions, all of which are
well-defined and gauge invariant. It is expressed as
\begin{eqnarray}\label{eq:factorization}
C(t)\otimes H(x,t)\otimes
\Phi(x)\otimes\exp[-s(P,b)-2\int^t_{1/b}\frac{d\mu}{\mu}\gamma_q(\alpha_s(\mu))],
\end{eqnarray}
where $C(t)$ are the corresponding Wilson coefficients of effective
operators defined in eq.(\ref{eq:H1},\ref{eq:H2}). Since the
transverse momentum of quark is kept in the pQCD approach, the large
double logarithm  $\ln^2 (Pb)$  (with P denoting the longitudinal
momentum, and b the conjugate variable of the transverse momentum)
to spoil the perturbative expansion. A resummation is thus needed to
give a
 Sudakov  factor $\exp[-s(P, b)]$ \cite{npb193381}. The term after Sudakov
 is from renormalization group running with  $\gamma_q=-\alpha_s/\pi$   the
quark anomalous dimension in axial gauge and $t$ the factorization
scale. All non-perturbative components are organized in the form of
hadron wave functions $\Phi(x)$ (with x the longitudinal momentum
fraction of valence quark inside the meson), which can be extracted
from experimental data or other non-perturbative methods. Since the
universal non-perturbative dynamics has been factored out, one can
evaluate all possible Feynman diagrams for the hard subamplitude
$H(x,t)$ straightforwardly, which include both traditional
factorizable and so-called ``non-factorizable" contributions.
Factorizable and non-factorizable annihilation type diagrams are
also calculable without end-point singularity.

\subsection{  Channels with both emission and annihilation contributions}

\begin{figure}[!htbh]
\begin{center}
\vspace{-2cm} \centerline{\epsfxsize=12 cm \epsffile{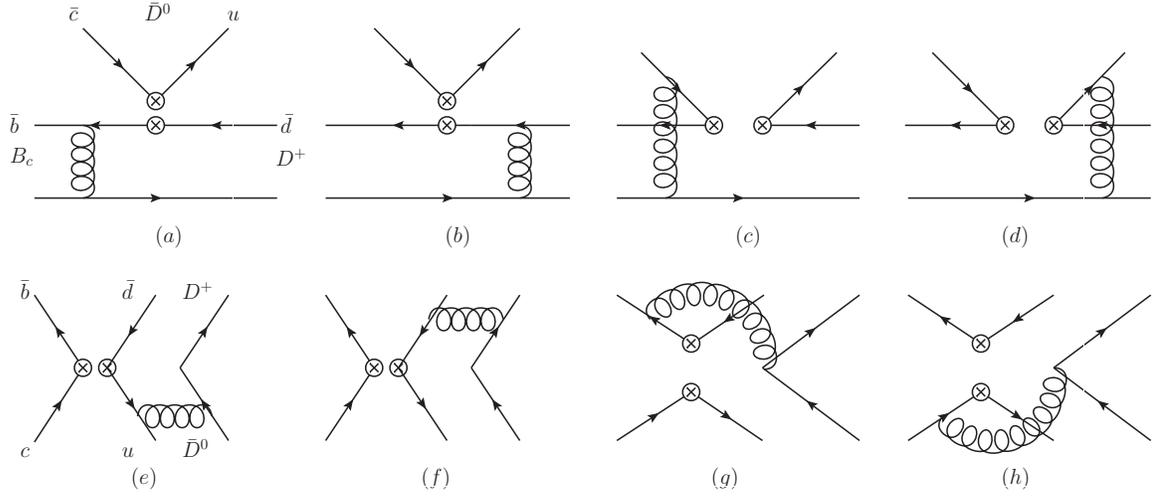}}
\vspace{-8.4cm} \caption{Feynman diagrams for  $B_c\rightarrow
D^+\bar{D}^0$ decays.}
 \label{fig:bctod0b}
 \end{center}
\end{figure}

At leading order, there are eight kinds of  Feynman diagrams
contributing to this type of CKM favored decays according to
eq.(\ref{eq:H1}). Here, we take the decay $B_c\rightarrow
D^+\bar{D}^0$ as an example, whose Feynman diagrams
 are shown in Fig.\ref{fig:bctod0b}.
The first line are the emission  type diagrams, with the first two
contributing to the usual form factor; the last two so-called
``non-factorizable" diagrams. In fact, the first two diagrams are
the only contributions calculated in the naive factorization
approach. The second line are the annihilation type diagrams, with
the first two factorizable; the last two non-factorizable. The decay
amplitude of factorizable diagrams (a) and (b) in
Fig.\ref{fig:bctod0b} is
\begin{eqnarray}\label{eq:fe}
\mathcal {F}_e&=&-2\sqrt{\frac{2}{3}}C_Ff_Bf_{3}\pi M_B^4
\int_0^1dx_2\int_0^{\infty}b_1b_2db_1db_2\phi_{2}(x_2)\exp(-\frac{b_1^2\omega_B^2}{2})\times\nonumber\\&&\{
[-(r_2-2)r_b+2r_2x_2-x_2]\alpha_s(t_a)h_e(\alpha_e,\beta_a,b_1,b_2)S_t(x_2)\exp[-S_{ab}(t_a)]
\nonumber\\&&+2r_2\alpha_s(t_b)h_e(\alpha_e,\beta_b,b_2,b_1)S_t(x_1)\exp[-S_{ab}(t_b)]\},
\end{eqnarray}
where  $r_b=m_b/M_B$, $r_i=m_i/M_B(i=2,3)$ with $m_2,m_3$ are the masses of
 the recoiling charmed meson and the emitting charmed meson, respectively;
 $C_F=4/3$ is a color factor; $f_3$ is the decay constant of the
charmed meson, which emitted from the weak vertex. The factorization
scales $t_{a,b}$ are chosen as the maximal virtuality of internal
particles in the hard amplitude, in order to suppress the higher
order corrections \cite{prd074004}. The function $h_e$
 and the Sudakov factor $\exp[-S]$ are
displayed in the Appendix \ref{sec:b}. $D$ meson distribution amplitude $\phi(x)$ are given in Appendix \ref{sec:c}.
 The factor $S_t(x)$ is the jet function resulting from the threshold resummation,
 whose definitions can be found in \cite{epjc45711}.

The formula for non-factorizable emission diagrams Fig.
\ref{fig:bctod0b} (c) and (d) contain the kinematics variables of
all the three mesons. Its expression is:
\begin{eqnarray}\label{eq:me}
\mathcal {M}_e&=&-\frac{8}{3}C_Ff_B\pi M_B^4
\int_0^1dx_2dx_3\int_0^{\infty}b_2b_3db_2db_3\phi_{2}(x_2)\phi_{3}(x_3)\exp(-\frac{b_2^2\omega_B^2}{2})\times\nonumber\\&&
\{[1-x_1-x_3-r_2(1-x_2)]\alpha_s(t_c)h_e(\beta_c,\alpha_e,b_3,b_2)\exp[-S_{cd}(t_c)]-
\nonumber\\&&[1-x_1-x_2+x_3-r_2(1-x_2)]\alpha_s(t_d)h_e(\beta_d,\alpha_e,b_3,b_2)\exp[-S_{cd}(t_d)]
\}.
\end{eqnarray}
Generally, for charmless decays of B meson, the non-factorizable
contributions of the emission diagrams are   small due to the
cancelation between Fig. \ref{fig:bctod0b} (c) and (d). While for
double charm decays with the light meson replaced by a charmed
meson, since the heavy $\bar{c}$ quark and the light quark is not
symmetric, the non-factorizable emission diagrams  ought to give
remarkable contributions. This has been shown in the pQCD
calculation of $B\to D\pi$ decays for a very large branching ratios
of color-suppressed modes \cite{dpi} and proved by the B factory
experiments.

The    decay amplitude of  factorizable annihilation diagrams Fig.
\ref{fig:bctod0b} (e) and (f) involve only the two final states
charmed  meson wave functions, shown as
\begin{eqnarray}
\mathcal {F}_a&=&-8C_Ff_B\pi M_B^4
\int_0^1dx_2dx_3\int_0^{\infty}b_2b_3db_2db_3\phi_{2}(x_2)\phi_{3}(x_3)\times\nonumber\\&&
\{[1-x_2]\alpha_s(t_e)h_e(\alpha_a,\beta_e,b_2,b_3)\exp[-S_{ef}(t_e)]S_t(x_3)-
\nonumber\\&&[1-x_3]
\alpha_s(t_f)h_e(\alpha_a,\beta_f,b_3,b_2)\exp[-S_{ef}(t_f)]S_t(x_2)
\}.
\end{eqnarray}
For the non-factorizable annihilation diagrams Fig.
\ref{fig:bctod0b} (g) and (h), the decay amplitude is
\begin{eqnarray}
\mathcal {M}_a&=&\frac{8}{3}C_Ff_B\pi M_B^4
\int_0^1dx_2dx_3\int_0^{\infty}b_1b_2db_1db_2\phi_{2}(x_2)\phi_{3}(x_3)\exp(-\frac{b_1^2\omega_B^2}{2})\nonumber\\&&\times
\{[x_1+x_3-1-r_c]\alpha_s(t_g)h_e(\beta_g,\alpha_a,b_1,b_2)\exp[-S_{gh}(t_g)]\nonumber\\&&-
[r_b-x_2]\alpha_s(t_h)h_e(\beta_h,\alpha_a,b_1,b_2)\exp[-S_{gh}(t_h)]
\},
\end{eqnarray}
where $r_c=m_c/M_B$, with $m_c$   the mass of c quark in $B_c$
meson. Finally, the total decay amplitude for  $B_c\rightarrow
D^+\bar{D}^0$  can be given by
\begin{eqnarray}\label{eq:dpd0b}
\mathcal {A}(B_c\rightarrow D^+\bar{D}^0)&=&V_{cb}^*V_{ud}[a_2\mathcal {F}_e+C_2\mathcal {M}_e+a_1\mathcal {F}_a+C_1\mathcal {M}_a],
\end{eqnarray}
 with the combinations of Wilson coefficients $a_1=C_2+C_1/3$ and
 $a_2=C_1+C_2/3$, characterizing the color favored contribution
 and  the color-suppressed contribution in the naive factorization, respectively.
The total decay amplitudes of $B_c\rightarrow D_s^+\bar{D}^0$, $B_c\rightarrow D^+\bar{D}^{*0}$  and $B_c\rightarrow D_s^+\bar{D}^{*0}$
can be  obtained from eq.(\ref{eq:dpd0b})
with the following  replacement:
\begin{eqnarray}\label{eq:relation1}
\mathcal {A}(B_c\rightarrow D_s^+\bar{D}^0)&=&V_{cb}^*V_{us}[a_2\mathcal {F}_e+C_2\mathcal {M}_e+a_1\mathcal {F}_a+C_1\mathcal {M}_a]|
_{ D^+\rightarrow D^+_s },\nonumber\\
\mathcal {A}(B_c\rightarrow D^+\bar{D}^{*0})&=&V_{cb}^*V_{ud}[a_2\mathcal {F}_e+C_2\mathcal {M}_e+a_1\mathcal {F}_a+C_1\mathcal {M}_a]
|_{\bar{D}^0\rightarrow \bar{D}^{*0} },\nonumber\\
\mathcal {A}(B_c\rightarrow D_s^+\bar{D}^{*0})&=&V_{cb}^*V_{us}[a_2\mathcal {F}_e+C_2\mathcal {M}_e+a_1\mathcal {F}_a+C_1\mathcal {M}_a]
|_{ D^+\rightarrow D^+_s,\bar{D}^0\rightarrow \bar{D}^{*0}  }.
\end{eqnarray}
Comparing our eq.(\ref{eq:dpd0b},\ref{eq:relation1}) with the
formulas of previous naive factorization approach
\cite{jpg301445,plb555189,prd73054024,prd564133,pan67,prd61034012},
it is easy to see that only the first term appearing in
eq.(\ref{eq:dpd0b},\ref{eq:relation1}) are calculated in the
previous naive factorization approach. The second, third and fourth
terms in these equations, are the corresponding non-factorizable
emission type contribution, factorizable and non-factorizable
annihilation type contributions, respectively, which are all new
calculations.

In $B_c\rightarrow D^{*+}_{(s)}\bar{D}^{*0}$ decays, the two vector
mesons in the final states have the same helicity due to angular
momentum conservation, therefore only  three different polarization
states, one longitudinal and two transverse for both vector mesons,
are possible. The decay amplitude can be decomposed as
\begin{eqnarray}
\mathcal {A}=\mathcal {A}^L+\mathcal {A}^N \epsilon_{2}^T \cdot \epsilon_{3}^T +
i \mathcal {A}^T \epsilon_{\alpha\beta\rho\sigma}n^{\alpha}v^{\beta}\epsilon_{2}^{T \rho}\epsilon_{3}^{T \sigma},
\end{eqnarray}
where  $\epsilon_2^T, \epsilon_3^T$ are the transverse polarization
vectors for the two vector charmed mesons, respectively. $\mathcal
{A}^L$ corresponds to the contributions of  longitudinal
polarization; $\mathcal {A}^N$ and $\mathcal {A}^T$ corresponds to
the contributions of  normal  and transverse  polarization,
respectively. And the total amplitudes $\mathcal {A}^{L,N,T}$ have
the same structures as eq.(\ref{eq:dpd0b},\ref{eq:relation1}). The
factorization formulae for the longitudinal, normal and transverse
polarizations  are   listed in Appendix \ref{sec:a}.

For $B_c\rightarrow D^{*+}_{(s)}\bar{D}^0$ decays, only the
longitudinal polarization of $ D^{*+}_{(s)}$ meson will contribute,
due to the angular momentum conservation. We can obtain their decay
amplitudes from the longitudinal polarization amplitudes for the
$B_c\rightarrow D^{*+}_{(s)}\bar{D}^{*0}$ decays with the
replacement $\bar{D}^{*0}\rightarrow \bar{D}^{0}$.

\subsection{ Channels with  pure emission type decays}

\begin{figure}[t]
\begin{center}
 \vspace{-2cm}\centerline{\epsfxsize=12 cm \epsffile{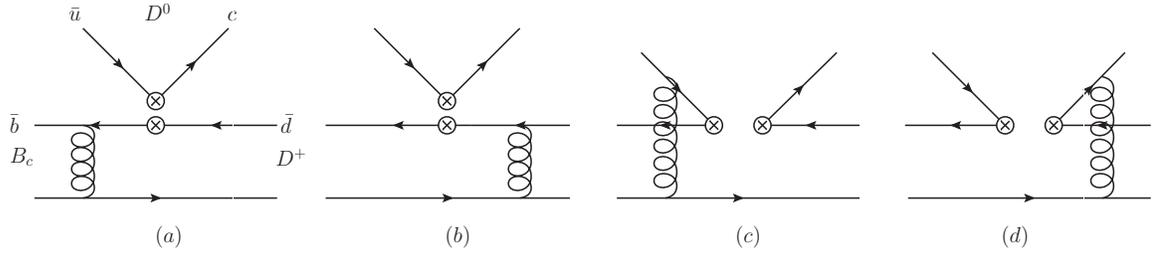}}
 \vspace{-12cm}\caption{Color-suppressed emission diagrams contributing to the $B_c\rightarrow D^+D^0$ decays.} \label{fig:bctod01}
 \end{center}
\end{figure}
\begin{figure}[t]
\begin{center}
 \vspace{-2cm}\centerline{\epsfxsize=12 cm \epsffile{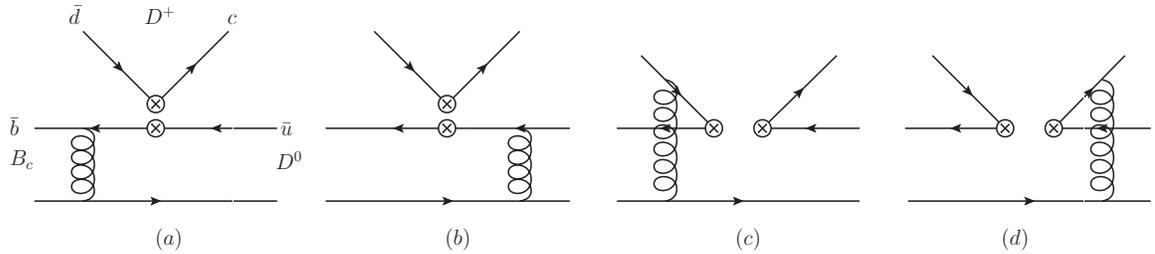}}
 \vspace{-12cm}\caption{Color-favored emission diagrams contributing to the $B_c\rightarrow D^+D^0$ decays.} \label{fig:bctod02}
 \end{center}
\end{figure}

There are also eight kinds of Feynman diagrams contributing to
$B_c\rightarrow D_{(s)}^{(*)+}D^{(*)0}$  decays according to
eq.(\ref{eq:H2}), but all are emission type. Taking the decay
$B_c\rightarrow D^+D^0$ as an example,
 Fig. \ref{fig:bctod01} are the color-suppressed emission diagrams while  Fig. \ref{fig:bctod02} are the color-favored emission diagrams.
We mark the subscript 2 and 3 to denote the contributions from Fig.
\ref{fig:bctod01} and Fig. \ref{fig:bctod02}, respectively. The
decay amplitude of factorization emission diagrams $\mathcal
{F}_{e2}$,   coming from  Fig. \ref{fig:bctod01} (a,b), is similar
to eq.(\ref{eq:fe}), but with the replacement $\bar{D}^0\rightarrow
D^0$.
 While the decay amplitude of non-factorization emission diagram $\mathcal {M}_{e2}$,
  coming from  Fig. \ref{fig:bctod01} (c,d), is different from  eq.(\ref{eq:me}), since
the heavy c quark and the light anti-quark are not symmetric. The
expression of the non-factorizable emission diagram is
\begin{eqnarray}\label{eq:mep}
\mathcal {M}_{e2}&=&-\frac{8}{3}C_Ff_B\pi M_B^4
\int_0^1dx_2dx_3\int_0^{\infty}b_2b_3db_2db_3\phi_{2}(x_2)\phi_{3}(x_3)\exp(-\frac{b_2^2\omega_B^2}{2})\nonumber\\&&\times
\{[2-x_1-x_2-x_3-r_2(1-x_2)]\alpha_s(t_c)h_e(\beta_c,\alpha_e,b_3,b_2)\exp[-S_{cd}(t_c)]-
\nonumber\\&&[x_3-x_1-r_2(1-x_2)]\alpha_s(t_d)h_e(\beta_d,\alpha_e,b_3,b_2)\exp[-S_{cd}(t_d)]
\}.
\end{eqnarray}
By exchanging the two final states charmed mesons in
Fig.~\ref{fig:bctod01}, one can obtain the  corresponding decay
amplitudes formulae $\mathcal {F}_{e3}$ and $\mathcal {M}_{e3}$ for
Fig.~\ref{fig:bctod02}.
 The total decay amplitude of $B_c\rightarrow D^+D^0$ decay can be written as
\begin{eqnarray}\label{eq:bctodpd0}
\mathcal {A}(B_c\rightarrow D^+D^0)&=&V_{ub}^*V_{cd}[a_2\mathcal {F}_{e2}+C_2\mathcal {M}_{e2}
+a_1\mathcal {F}_{e3}+C_1\mathcal {M}_{e3}].
\end{eqnarray}
If the final recoiling meson is the vector $D^*$ meson, the decay
amplitudes of factorization emission diagrams and non-factorization
emission diagrams are given as
\begin{eqnarray}\label{eq:dstarfe}
\mathcal {F}^*_{e2}&=&-2\sqrt{\frac{2}{3}}C_Ff_Bf_{3}\pi M_B^4
\int_0^1dx_2\int_0^{\infty}b_1b_2db_1db_2\phi_{2}(x_2)\exp(-\frac{b_1^2\omega_B^2}{2})\nonumber\\&&\times\{
[-(r_2-2)r_b+2r_2x_2-x_2]\alpha_s(t_a)h_e(\alpha_e,\beta_a,b_1,b_2)S_t(x_2)\exp[-S_{ab}(t_a)]
\nonumber\\&&+r^2_2\alpha_s(t_b)h_e(\alpha_e,\beta_b,b_2,b_1)S_t(x_1)\exp[-S_{ab}(t_b)]\},
\end{eqnarray}
\begin{eqnarray}
\mathcal {M}^*_{e2}&=&-\frac{8}{3}C_Ff_B\pi M_B^4
\int_0^1dx_2dx_3\int_0^{\infty}b_2b_3db_2db_3\phi_{2}(x_2)\phi_{3}(x_3)\exp(-\frac{b_2^2\omega_B^2}{2})\nonumber\\&&\times
\{[2-x_1-x_2-x_3-r_2(1-x_2)]\alpha_s(t_c)h_e(\beta_c,\alpha_e,b_3,b_2)\exp[-S_{cd}(t_c)]-
\nonumber\\&&[x_3-x_1+r_2(1-x_2)]\alpha_s(t_d)h_e(\beta_d,\alpha_e,b_3,b_2)\exp[-S_{cd}(t_d)]
\}.\nonumber\\&&
\end{eqnarray}
The total decay amplitudes  for other  pure emission type decays are then
\begin{eqnarray}\label{eq:bctod0}
\mathcal {A}(B_c\rightarrow D_s^+D^0)&=&V_{ub}^*V_{cs}[a_2\mathcal {F}_{e2}+C_2\mathcal {M}_{e2}
+a_1\mathcal {F}_{e3}+C_1\mathcal {M}_{e3}],\nonumber\\
\mathcal {A}(B_c\rightarrow D^+D^{*0})&=&V_{ub}^*V_{cd}[a_2\mathcal {F}_{e2}+C_2\mathcal {M}_{e2}
+a_1\mathcal {F}^*_{e3}+C_1\mathcal {M}^*_{e3}],\nonumber\\
\mathcal {A}(B_c\rightarrow D^{*+}D^{0})&=&V_{ub}^*V_{cd}[a_2\mathcal {F}^*_{e2}+C_2\mathcal {M}^*_{e2}
+a_1\mathcal {F}_{e3}+C_1\mathcal {M}_{e3}],\nonumber\\
\mathcal {A}(B_c\rightarrow D_s^+D^{*0})&=&V_{ub}^*V_{cs}[a_2\mathcal {F}_{e2}+C_2\mathcal {M}_{e2}
+a_1\mathcal {F}^*_{e3}+C_1\mathcal {M}^*_{e3}],\nonumber\\
\mathcal {A}(B_c\rightarrow D_s^{*+}D^{0})&=&V_{ub}^*V_{cs}[a_2\mathcal {F}^*_{e2}+C_2\mathcal {M}^*_{e2}
+a_1\mathcal {F}_{e3}+C_1\mathcal {M}_{e3}].\nonumber\\
\end{eqnarray}
The $B_c\rightarrow D^{*+}_{(s)}D^{*0}$ decays have a similar
situation to  $B_c\rightarrow D^{*+}_{(s)}\bar{D}^{*0}$,
their factorization formulae are also listed in Appendix.\ref{sec:a}.

\section{NUMERICAL RESULTS} \label{sec:result}

In this section, we summarize the numerical results and analysis in
the double charm decays of the $B_c$ meson. Some input parameters
needed in the pQCD calculation   are listed in Table
\ref{tab:constant}.

\subsection{The Form Factors}
\begin{table}
\caption{Parameters we used in numerical calculation \cite{npp37}}
\label{tab:constant}
\begin{tabular*}{13.5cm}{@{\extracolsep{\fill}}l|cccc}
  \hline\hline
\textbf{Mass(\text{GeV})} & $M_{W}=80.399$   & $M_{B_c}=6.277$ & $m_{b}=4.2$& $m_{c}=1.27$\\[1ex]
\hline
\end{tabular*}
\begin{tabular*}{13.5cm}{@{\extracolsep{\fill}}l|ccc}
  \hline
\multirow{2}{*}{{\textbf{CKM}}} & $|V_{ub}|=(3.47^{+0.16}_{-0.12})\times 10^{-3}$
 & $|V_{ud}|=0.97428^{+0.00015}_{-0.00015}$ & $|V_{us}|=0.2253^{+0.0007}_{-0.0007}$\\[1ex]
&$|V_{cs}|=0.97345^{+0.00015}_{-0.00016}$ & $|V_{cd}|=0.2252^{+0.0007}_{-0.0007}$
& $|V_{cb}|=0.0410^{+0.0011}_{-0.0007}$  \\[1ex]
\hline
\end{tabular*}
\begin{tabular*}{13.5cm}{@{\extracolsep{\fill}}l|ccc}
\hline
\textbf{Decay constants(MeV)} & $f_{B_c}=489$ & $f_{D}=206.7 \pm 8.9$ & $f_{D_s}=257.5 \pm 6.1$\\[1ex]
\hline
\end{tabular*}
\begin{tabular*}{13.5cm}{@{\extracolsep{\fill}}l|l}
\hline
\textbf{Lifetime} & $\tau_{B_c}=0.453\times 10^{-12}\text{s}$\\[1ex]
\hline\hline
\end{tabular*}
\end{table}

\begin{table}
\caption{The form factors for $B_c\rightarrow D^{(*)}_{(s)}$ at
$q^2=0$ evaluated in the pQCD approach. The  uncertainties are from
the hadronic parameters. For comparison, we also cite the
theoretical estimates of other models.} \label{tab:formfactor}
\begin{tabular}[t]{l|c|c|c|c|c|c}
\hline\hline
 & This work & Kiselev \cite{jpg301445} \footnotemark[1]& IKP \cite{plb555189} & WSL \cite{prd7905402}
 & DSV \cite{jpg35085002}& DW \cite{prd391342}\footnotemark[2] \\ \hline
$F^{B_c\rightarrow D}$       & $0.14^{+0.01}_{-0.02}$  &0.32 [0.29]  &0.189   &0.16 &0.075&0.255\\
$F^{B_c\rightarrow D_s}$     & $0.19^{+0.02}_{-0.01}$ &0.45 [0.43]    &0.194 &0.28 &0.15  &--\\
$A_0^{B_c\rightarrow D^*}$   & $0.12^{+0.02}_{-0.01}$  &0.35 [0.37]   &0.133  &0.09 &0.081 &0.257\\
$A_0^{B_c\rightarrow D_s^*}$ & $0.17^{+0.01}_{-0.01}$ &0.47 [0.52]    &0.142  &0.17 &0.16  &--\\
\hline\hline
\end{tabular}
\footnotetext[1]{ The non-bracket (bracketed) results are evaluated
in sum rules (potential model)} \footnotetext[2]{We quote  the
result with $\omega=0.7\text{GeV}$}
\end{table}

The diagrams (a) and (b) in Fig.\ref{fig:bctod0b} or
Fig.\ref{fig:bctod02} give the contribution for $B_c\rightarrow
D^{(*)}_{(s)}$ transition form factor at  $q^2=0$ point. Our
predictions of the   form factors are collected in Table
\ref{tab:formfactor}. The  error  is from the combined uncertainty
in  the hadronic parameters: (1) the shape parameters:
$\omega_B=0.60\pm0.05$ for $B_c$ meson wave function, $a_{D}=(0.5\pm
0.1) \text{GeV}$ for $D^{(*)}$ meson and $a_{D_s}=(0.4\pm 0.1)
\text{GeV}$ for $D_s^{(*)}$ meson wave function \cite{dd1}; (2) the
decay constants  in the wave functions of charmed mesons, which are
given in  Table \ref{tab:constant}. Since the uncertainties from
decay constants of $D_{(s)}$ and  the shape parameters of the wave
functions  are very small, the  relevant uncertainties to the form
factors are also very small. We can see that the $SU(3)$ symmetry
breaking effects between $B_c$ to $ D^{(*)}$ and $B_c$ to
$D^{(*)}_{s}$ form factors are   large, as the decay constant of
$D_s$ is about one-fifth larger than that of the $D$ meson.

In the literature there are already   lots of studies on
$B_c\rightarrow D^{(*)}_{(s)}$ transition form factors
\cite{jpg301445,prd7905402,jpg35085002,plb555189,prd391342}, whose
  results are collected in Table \ref{tab:formfactor}. Our
results are generally close to the covariant light-front quark model
results of \cite{prd7905402} and the constituent quark model results
of \cite{plb555189}. However, other results collected in Table
\ref{tab:formfactor}, especially for the QCD sum rules (QCDSR)
\cite{jpg301445} and the Bauer, Stech and Wirbel (BSW) model
\cite{jpg35085002} deviate a lot numerically. The predictions of
QCDSR \cite{jpg301445} are larger than those in other works
\cite{prd7905402,jpg35085002,plb555189,prd391342}. The reason is
that they have taken into account the $\alpha_s/v$ corrections and
the form factors are enhanced by 3 times due to the Coulomb
renormalization of the quark-meson vertex for the heavy quarkonium
$B_c$. The results of BSW model \cite{jpg35085002} are quite small
due to the less overlap of the initial and final states wave
functions. Although, the included flavor dependence of the average
transverse quark momentum in the mesons can enhance the form factors
for $B_c\rightarrow D^{*}_{(s)}$ transitions, their predictions are
still smaller than  other models. The large differences in different
models   can be discriminated by the future LHC experiments.

\subsection{Branching Ratios}

With the decays amplitudes $\mathcal {A}$ obtained in
Sec.\ref{sec:f-work}, the branching ratio $\mathcal {BR}$ reads as
\begin{eqnarray}
\mathcal {BR}=\frac{G_F\tau_{B_c}}{32\pi M_B}\sqrt{(1-(r_2+r_3)^2)(1-(r_2-r_3)^2)}|\mathcal {A}|^2.
\end{eqnarray}
As stated in Sec \ref{sec:f-work}, the contributions from the
penguin operators are absent, since the penguins add an even number
of charmed quarks,  while there is already  one from the initial
state. There should be no CP violation in these processes. We
tabulate the branching ratios  of the  considered decays  in Table
\ref{tab:branching1} and \ref{tab:branching2}. The processes (1)-(4)
in  Table \ref{tab:branching1} have a comparatively large branching
ratios ($10^{-5}$)
  with the CKM
 factor $V_{cb}^*V_{ud}\sim \lambda^2$. While the branching ratios of other processes are relatively small
 due to the CKM factor suppression. Especially for the processes (1)-(4) in Table \ref{tab:branching2},
 these channels are suppressed by CKM element $V_{ub}/V_{cb}$ and  $V_{cd}/V_{ud}$.
 Thus their branching ratios are three order magnitudes smaller.

\begin{table}
\caption {Branching ratios ($10^{-6}$) of the CKM favored decays
with both emission and annihilation contributions, together with
results from other models. The errors for these entries correspond
to the uncertainties in the input hadronic quantities, from the CKM
matrix elements, and the scale dependence, respectively.}
\label{tab:branching1}
\begin{tabular}[t]{l|l|c|ccccc}
\hline\hline
 &channels & This work & Kiselev\cite{jpg301445} & IKP\cite{plb555189} & IKS\cite{prd73054024}
 &LC\cite{prd564133} & CF\cite{prd61034012} \\ \hline
1&$B_c\rightarrow D^+\bar{D}^0$       & $32^{+6+1+2}_{-6-1-4}$&53  &32&33 &86 &8.4             \\
2&$B_c\rightarrow D^+\bar{D}^{*0}$     &$34^{+7+2+3}_{-6-1-3}$&75  &83 &38  &75&7.5           \\
3&$B_c\rightarrow D^{*+}\bar{D}^0$   & $12^{+3+1+0}_{-3-0-1}$  &49 &17  &9   &30&84            \\
4&$B_c\rightarrow D^{*+}\bar{D}^{*0}$ &$34^{+9+2+0}_{-8-1-0}$  &330 &84&21&55 &140        \\
\hline
5&$B_c\rightarrow D_s^+\bar{D}^0$       & $2.3^{+0.4+0.1+0.2}_{-0.4-0.1-0.2}$  &4.8  &1.7&2.1&4.6  &0.6        \\
6&$B_c\rightarrow D_s^+\bar{D}^{*0}$     &$2.6^{+0.4+0.1+0.1}_{-0.6-0.1-0.2}$  &7.1  &4.3&2.4   &3.9 &0.53     \\
7&$B_c\rightarrow D_s^{*+}\bar{D}^0$   & $0.7^{+0.1+0.0+0.0}_{-0.2-0.0-0.0}$  &4.5&0.95  &0.65  &1.8 &5        \\
8&$B_c\rightarrow D_s^{*+}\bar{D}^{*0}$ & $2.8^{+0.7+0.1+0.1}_{-0.6-0.1-0.0}$ &26 &4.7 &1.6  &3.5&8.4      \\
\hline\hline
\end{tabular}
\end{table}

\begin{table}
\caption{Branching ratios ($10^{-7}$) of the CKM suppressed decays
with pure emission  contributions, together with results from other
models. The errors for these entries correspond to the uncertainties
in the input hadronic quantities, from the CKM matrix elements, and
the scale dependence, respectively.} \label{tab:branching2}
\begin{tabular}[t]{l|l|c|ccc}
\hline\hline
 &channels & This work & Kiselev\cite{jpg301445}  &  IKP\cite{plb555189}& IKS\cite{prd73054024}
  \\ \hline
1&$B_c\rightarrow D^+D^0$       & $1.0^{+0.2+0.1+0.0}_{-0.1-0.0-0.0}$  &3.2 &1.1&3.1 \\
2&$B_c\rightarrow D^+D^{*0}$     &$0.7^{+0.1+0.1+0.0}_{-0.2-0.0-0.0}$&2.8 &0.25&0.52 \\
3&$B_c\rightarrow D^{*+}D^0$   & $0.9^{+0.1+0.1+0.0}_{-0.2-0.0-0.0}$ &4.0&3.8&4.4\\
4&$B_c\rightarrow D^{*+}D^{*0}$ & $0.8^{+0.2+0.1+0.2}_{-0.1-0.0-0.0}$ &15.9&2.8&2.0 \\
\hline
5&$B_c\rightarrow D_s^+D^0$       & $30^{+5+3+1}_{-4-2-1}$  &66 &25&74 \\
6&$B_c\rightarrow D_s^+D^{*0}$     &$19^{+3+2+0}_{-3-1-1}$ &63 &6&13 \\
7&$B_c\rightarrow D_s^{*+}D^0$   & $25^{+4+2+0}_{-3-2-1}$  &85 &69&93\\
8&$B_c\rightarrow D_s^{*+}D^{*0}$ & $24^{+3+2+1}_{-3-2-1}$ &404&54&45 \\
\hline\hline
\end{tabular}
\end{table}

For comparison, we also cite other theoretical results
\cite{jpg301445,plb555189,prd73054024,prd564133,prd61034012} for the
double charm decays of $B_c$ meson in Tables \ref{tab:branching1}
and \ref{tab:branching2}.
In general, the results of the various model calculations  are of
the same order of magnitude for most channels. However the
difference between different model calculations is quite large. This
is expected from the large difference of input parameters,
especially the large difference of form factors shown in
Table~\ref{tab:formfactor}. As stated in the introduction, all the
calculations of these $B_c$ to two D meson decays in the literature
use the same naive factorization approach. Their difference relies
only on the input form factors and decay constants. Therefore the
comparison of results with any of them is straightforward. Larger
branching ratios come always with the larger form factors.
 As stated in the previous subsection,
our results of form factors  are comparable with the relativistic
constituent quark model (RCQM) \cite{plb555189,prd73054024}, thus
our branching ratios in Table \ref{tab:branching1} are also
comparable with theirs except for the processes $B_c\rightarrow
D^{*+}\bar D^{*0}$ and $B_c\rightarrow D_s^{*+}\bar D^{*0}$. Due to
the sizable contributions of transverse polarization amplitudes, our
branching ratios are  larger than those in RCQM model, whose
transverse contribution is negligible.

Since all the previous calculations in the literature are model
calculations, it is difficult for them to give the theoretical error
estimations. In our pQCD approach, the factorization holds at the
leading order expansion of $m_D/m_B$. At this order, we can do the
systematical calculation, so as to the error estimations in the
tables. The first error in these entries is estimated from the
hadronic parameters: (1) the shape parameters:
$\omega_B=0.60\pm0.05$ for $B_c$ meson, $a_{D}=(0.5\pm 0.1)
\text{GeV}$ for $D^{(*)}$ meson and $a_{D_s}=(0.4\pm 0.1)
\text{GeV}$ for $D_s^{(*)}$ meson \cite{dd1}; (2) the decay
constants in the wave functions of charmed mesons, which are given
in  Table \ref{tab:constant}.
The second error is from the uncertainty in the CKM matrix elements,
which are also given in Table \ref{tab:constant}. The third error
arises from the hard scale t varying from $0.75t$ to $1.25t$, which
characterizing the size of next-to-leading order QCD contributions.
The not large errors of this type indicate that our perturbative
expansion indeed hold. It is easy to see that the most important
uncertainty in our approach comes from the hadronic parameters. The
total theoretical error is in general around 10\% to 30\% in size.

The eight CKM favored channels (proportional to $|V_{cb}|$) in Table
\ref{tab:branching1} receive contributions from both emission
diagrams and annihilation diagrams. From Fig.\ref{fig:bctod0b}, one
can find that the contributions from the factorizable emission
diagrams are color-suppressed. The naive factorization approach can
not give reliable predictions due to large non-factorizable
contributions \cite{fac}. As was pointed out in
Sec.\ref{sec:f-work}, the non-factorizable emission diagrams give
large contributions in pQCD approach because the asymmetry of the
two quarks in charmed mesons.
 Thus, the branching ratios of these decays are dominated by the non-factorizable emission diagrams.

The eight CKM suppressed channels  (proportional to $|V_{ub}|$) in
Table \ref{tab:branching2} can occur only via emission type
diagrams. There are two types of emission diagrams in these decays,
one is color-suppressed, one is color favored. It is expected that
the color-favored factorizable amplitude $\mathcal {F}_{e3}$
dominates in eq.(\ref{eq:bctod0}). However, the non-factorizable
contribution $\mathcal {M}_{e2}$, proportional to the large $C_2$,
is enhanced by the Wilson coefficient. Numerically it is indeed
comparable to the color-favored factorizable amplitude. This large
non-factorizable contribution has already been shown in the similar
$B\to D\pi$ decays theoretically and experimentally \cite{dpi}. In
all of these channels the non-factorizable contributions play a very
important role, therefore the branching ratios predicted in
table~\ref{tab:branching1} and \ref{tab:branching2} are not  like
the previous naive factorization approach calculations
\cite{jpg301445,plb555189,prd73054024,prd564133,prd61034012}. They
are not simply proportional to the corresponding form factors any
more, but with a very complicated manner, since we have also
additional annihilation type contributions.

From Table III and IV, one can see that as it was expected the
magnitudes of the branching ratios of the decays $B_c\rightarrow
D^+_s \bar{D}^0$ and $B_c\rightarrow D^+_s D^0$ are very close to
each other. In our numerical results, the ratio of the two decay
widths is estimated as $\frac{\Gamma(B_c\rightarrow
D_s^+D^0)}{\Gamma(B_c\rightarrow D_s^+\bar{D}^0)}\approx 1.3$. They
are very suitable for extracting the CKM
 angle $\gamma$ though the amplitude relations. Hopefully   they will be measured in the
  experiments soon.
 However, the decays $B_c\rightarrow D^+ \bar{D}^0, D^+D^0$ are problematic from the methodic point of view for
  $\mathcal {BR}(B_c\rightarrow D^{+}D^0)\ll \mathcal {BR}(B_c\rightarrow D^{+}\bar{D}^0) $.
   The corresponding ratio in $B_c\rightarrow D^+D^0,
D^+\bar{D}^0$ decays is $\frac{\Gamma(B_c\rightarrow
D^+D^0)}{\Gamma(B_c\rightarrow D^+\bar{D}^0)} \sim 10^{-3}$, which
confirm the latter decay modes are not useful to determine the angle
$\gamma$  experimentally.

\begin{table}
\caption{The transverse polarizations fractions ($\%$) for
$B_c\rightarrow VV$.
 The errors correspond to the uncertainties in the hadronic parameters and the scale dependence, respectively. }
\label{tab:ratio}
\begin{tabular}[t]{l|c|c|c|c}
\hline\hline
 & $B_c\rightarrow D^{*+}\bar{D}^{*0}$ & $B_c\rightarrow D_s^{*+}\bar{D}^{*0}$
 & $B_c\rightarrow D^{*+}D^{*0}$ & $B_c\rightarrow D_s^{*+}D^{*0}$\\
\hline
$\mathcal {R}_T$ &$58^{+3+1}_{-3-0}$&$68^{+2+1}_{-2-1}$&$4^{+1+1}_{-1-1}$&$6^{+1+2}_{-0-1}$\\
\hline\hline
\end{tabular}
\end{table}

For the $B_c$ decays to two vector mesons, the decays amplitudes
$\mathcal {A}$ are defined in the helicity basis
\begin{eqnarray}
\mathcal {A}=\sum_{i=0,+,-}|\mathcal {A}_i|^2,\quad
\end{eqnarray}
where  the helicity amplitudes $\mathcal {A}_i$  have the following
relationships with $\mathcal {A}^{L,N,T}$
\begin{eqnarray}
\mathcal {A}_0=\mathcal {A}^L,\quad \mathcal {A}_{\pm}=\mathcal
{A}^N \pm \mathcal {A}^T.
\end{eqnarray}
We also calculate the transverse polarization fractions $\mathcal
{R}_T$ of the $B_c\to D_{(s)}^* D^*$ decays, with the definition
given by
\begin{eqnarray}
\mathcal {R}_T=\frac{|\mathcal {A}_+|^2+|\mathcal
{A}_-|^2}{|\mathcal {A}_0|^2+|\mathcal {A}_+|^2+|\mathcal {A}_-|^2}.
\end{eqnarray}
 This should be the first time
theoretical predictions in the literature, which are absent in all
the naive factorization calculations. According to the power
counting rules in the factorization assumption, the longitudinal
polarization should be dominant due to the quark helicity analysis.
Our predictions for the transverse polarization fractions of the
decays $B_c\rightarrow D^{*+}_{(s)}D^{*0}$, which are given in Table
\ref{tab:ratio}, are indeed small, since the two transverse
amplitudes are down by a power of $r_2$ or $r_3$ comparing with the
longitudinal amplitudes. However, for $B_c\rightarrow
D^{*+}_{(s)}\bar{D}^{*0}$ decays,
 the most
important contributions for these two decay channels are from the
non-factorizable tree diagrams in Fig. \ref{fig:bctod0b}(c) and
\ref{fig:bctod0b}(d). With an additional gluon, the transverse
polarization in the non-factorizable diagrams does not encounter
helicity flip suppression. The transverse polarization is at the
same order as longitudinal polarization. Therefore, we can expect
the transverse polarizations take a larger ratio in the  branching
ratios, which can reach $\sim 60\%$. The fact that the
non-factorizable contribution can give large transverse polarization
contribution is also observed in the $B^0 \to \rho^0\rho^0$,
$\omega\omega$ decays \cite{rhorho} and  in the $B_c\rightarrow
D_s^{*+}\omega$ decay \cite{11121257}.

\section{ conclusion}

All the previous calculations in the literature for the $B_c$ meson
decays to two charmed mesons are based on the very simple naive
factorization approach. The branching ratios predicted in this kind
of model calculation depend heavily on the input form factors. Since
all of these modes contain dominant or large contributions from
color-suppressed diagrams, the predicted branching ratios are also
not stable due to the large unknown non-factorizable contributions.
In this paper,
 we have performed a systematic
analysis of the double charm decays of the $B_c$ meson in the pQCD
approach based on $k_T$ factorization theorem,  which  is free of
end-point singularities.
 All topologies of decay amplitudes are
calculable in the same framework, including the  non-factorizable
one and annihilation type. It is found that the non-factorizable
emission diagrams give a remarkable contribution.
 There is no CP violation for all these decays within the standard model, since there
are only  tree operators contributions.
 The predicted branching ratios range from very small numbers of $\mathcal {O}(10^{-8})$ up to the
 largest branching fraction of $\mathcal {O}(10^{-5})$. 
Since all of the previous naive factorization calculations did  not
give the theoretical uncertainty in the numerical results, it is not
easy to compare our results with theirs. The theoretical uncertainty
study in the pQCD approach shows that our numerical results are
reliable, which may be tested in the upcoming experimental
measurements. We predict the transverse polarization fractions of
the $B_c$ decays with two vector $D^*$ mesons in the final states
for the first time. Due to the cancelation of some hadronic
parameters in the ratio, the polarization fractions are predicted
with less theoretical uncertainty. The transverse polarization
fractions are large in some channels, which mainly come from the
non-factorizable emission diagrams.

\begin{acknowledgments}

We thank  Hsiang-nan Li and Fusheng Yu for helpful
discussions. This work is partially supported by
 National Natural
Science Foundation of China under the Grant No.
11075168; Natural Science Foundation of Zhejiang Province of China,
  Grant No. Y606252 and Scientific Research Fund of Zhejiang Provincial Education Department of China, Grant No. 20051357.

\end{acknowledgments}

\begin{appendix}

\section{Factorization formulas for $B_c\rightarrow VV$}\label{sec:a}

In the $B_c$ decays to two vector meson final states, we use the
superscript L, N and T to denote the contributions from longitudinal
polarization, normal polarization and transverse polarization,
respectively. For the CKM favored $B_c\rightarrow
D^{*+}_{(s)}\bar{D}^{*0}$ decays, the decay amplitudes for different
polarizations are
\begin{eqnarray}\label{eq:fel}
\mathcal {F}^L_e&=&-2\sqrt{\frac{2}{3}}C_Ff_Bf_{3}\pi M_B^4
\int_0^1dx_2\int_0^{\infty}b_1b_2db_1db_2\phi_{2}(x_2)\exp(-\frac{b_1^2\omega_B^2}{2})\times\nonumber\\&&\{
[-(r_2-2)r_b+2r_2x_2-x_2]\alpha_s(t_a)h_e(\alpha_e,\beta_a,b_1,b_2)S_t(x_2)\exp[-S_{ab}(t_a)]
\nonumber\\&&+r^2_2\alpha_s(t_b)h_e(\alpha_e,\beta_b,b_2,b_1)S_t(x_1)\exp[-S_{ab}(t_b)]\},
\end{eqnarray}
\begin{eqnarray}\label{eq:mel}
\mathcal {M}^L_e&=&-\frac{8}{3}C_Ff_B\pi M_B^4
\int_0^1dx_2dx_3\int_0^{\infty}b_2b_3db_2db_3\phi_{2}(x_2)\phi_{3}(x_3)\exp(-\frac{b_2^2\omega_B^2}{2})\times\nonumber\\&&
\{[1-x_1-x_3+r_2(1-x_2)]\alpha_s(t_c)h_e(\beta_c,\alpha_e,b_3,b_2)\exp[-S_{cd}(t_c)]-
\nonumber\\&&[1-x_1-x_2+x_3-r_2(1-x_2)]\alpha_s(t_d)h_e(\beta_d,\alpha_e,b_3,b_2)\exp[-S_{cd}(t_d)]
\},
\end{eqnarray}
\begin{eqnarray}
\mathcal {F}^L_a&=&-8C_Ff_B\pi M_B^4
\int_0^1dx_2dx_3\int_0^{\infty}b_2b_3db_2db_3\phi_{2}(x_2)\phi_{3}(x_3)\times\nonumber\\&&
\{[1-x_2]\alpha_s(t_e)h_e(\alpha_a,\beta_e,b_2,b_3)\exp[-S_{ef}(t_e)]S_t(x_3)-
\nonumber\\&&[1-x_3]
\alpha_s(t_f)h_e(\alpha_a,\beta_f,b_3,b_2)\exp[-S_{ef}(t_f)]S_t(x_2)
\},
\end{eqnarray}
\begin{eqnarray}
\mathcal {M}^L_a&=&\frac{8}{3}C_Ff_B\pi M_B^4
\int_0^1dx_2dx_3\int_0^{\infty}b_1b_2db_1db_2\phi_{2}(x_2)\phi_{3}(x_3)\exp(-\frac{b_1^2\omega_B^2}{2})\times\nonumber\\&&
\{[x_1+x_3-1-r_c]\alpha_s(t_g)h_e(\beta_g,\alpha_a,b_1,b_2)\exp[-S_{gh}(t_g)]\nonumber\\&&-
[r_b-x_2]\alpha_s(t_h)h_e(\beta_h,\alpha_a,b_1,b_2)\exp[-S_{gh}(t_h)]
\},
\end{eqnarray}
\begin{eqnarray}\label{eq:fen}
\mathcal {F}^N_e&=&-2\sqrt{\frac{2}{3}}C_Ff_Bf_{3}r_3\pi M_B^4
\int_0^1dx_2\int_0^{\infty}b_1b_2db_1db_2\phi_{2}(x_2)\exp(-\frac{b_1^2\omega_B^2}{2})\times\nonumber\\&&\{
[2-r_b+r_2(4r_b-x_2-1)]\alpha_s(t_a)h_e(\alpha_e,\beta_a,b_1,b_2)S_t(x_2)\exp[-S_{ab}(t_a)]
\nonumber\\&&-r_2\alpha_s(t_b)h_e(\alpha_e,\beta_b,b_2,b_1)S_t(x_1)\exp[-S_{ab}(t_b)]\},
\end{eqnarray}
\begin{eqnarray}
\mathcal {F}^T_e&=&2\sqrt{\frac{2}{3}}C_Ff_Bf_{3}r_3\pi M_B^4
\int_0^1dx_2\int_0^{\infty}b_1b_2db_1db_2\phi_{2}(x_2)\exp(-\frac{b_1^2\omega_B^2}{2})\times\nonumber\\&&\{
[2-r_b-r_2(1-x_2)]\alpha_s(t_a)h_e(\alpha_e,\beta_a,b_1,b_2)S_t(x_2)\exp[-S_{ab}(t_a)]
\nonumber\\&&-r_2\alpha_s(t_b)h_e(\alpha_e,\beta_b,b_2,b_1)S_t(x_1)\exp[-S_{ab}(t_b)]\},
\end{eqnarray}
\begin{eqnarray}
\mathcal {M}^N_e&=&-\mathcal {M}^T_e=\frac{8}{3}C_Ff_B\pi M_B^4r_3
\int_0^1dx_2dx_3\int_0^{\infty}b_2b_3db_2db_3\phi_{2}(x_2)\phi_{3}(x_3)\exp(-\frac{b_2^2\omega_B^2}{2})\nonumber\\&&\times
\{[x_1+x_3-1]\alpha_s(t_c)h_e(\beta_c,\alpha_e,b_3,b_2)\exp[-S_{cd}(t_c)]-
\nonumber\\&&[x_1-x_3]\alpha_s(t_d)h_e(\beta_d,\alpha_e,b_3,b_2)\exp[-S_{cd}(t_d)]\},
\end{eqnarray}
\begin{eqnarray}
\mathcal {F}^N_a&=&-8C_Ff_B\pi M_B^4r_2r_3
\int_0^1dx_2dx_3\int_0^{\infty}b_2b_3db_2db_3\phi_{2}(x_2)\phi_{3}(x_3)\times\nonumber\\&&
\{[2-x_2]\alpha_s(t_e)h_e(\alpha_a,\beta_e,b_2,b_3)\exp[-S_{ef}(t_e)]S_t(x_3)-
\nonumber\\&&[2-x_3]
\alpha_s(t_f)h_e(\alpha_a,\beta_f,b_3,b_2)\exp[-S_{ef}(t_f)]S_t(x_2)\},
\end{eqnarray}
\begin{eqnarray}
\mathcal {F}^T_a&=&-8C_Ff_B\pi M_B^4r_2r_3
\int_0^1dx_2dx_3\int_0^{\infty}b_2b_3db_2db_3\phi_{2}(x_2)\phi_{3}(x_3)\times\nonumber\\&&
\{x_2\alpha_s(t_e)h_e(\alpha_a,\beta_e,b_2,b_3)\exp[-S_{ef}(t_e)]S_t(x_3)+
\nonumber\\&&x_3
\alpha_s(t_f)h_e(\alpha_a,\beta_f,b_3,b_2)\exp[-S_{ef}(t_f)]S_t(x_2)\},
\end{eqnarray}
\begin{eqnarray}
\mathcal {M}^N_a&=&\frac{8}{3}C_Ff_B\pi M_B^4
\int_0^1dx_2dx_3\int_0^{\infty}b_1b_2db_1db_2\phi_{2}(x_2)\phi_{3}(x_3)\exp(-\frac{b_1^2\omega_B^2}{2})\times\nonumber\\&&
\{[r_2^2(x_2-1)+r_3^2(x_3-1)]\alpha_s(t_g)h_e(\beta_g,\alpha_a,b_1,b_2)\exp[-S_{gh}(t_g)]\nonumber\\&&-
[r_2^2x_2+r_3^2x_3-2r_2r_3r_b]\alpha_s(t_h)h_e(\beta_h,\alpha_a,b_1,b_2)\exp[-S_{gh}(t_h)]\},
\end{eqnarray}
\begin{eqnarray}
\mathcal {M}^T_a&=&\frac{8}{3}C_Ff_B\pi M_B^4
\int_0^1dx_2dx_3\int_0^{\infty}b_1b_2db_1db_2\phi_{2}(x_2)\phi_{3}(x_3)\exp(-\frac{b_1^2\omega_B^2}{2})\times\nonumber\\&&
\{[r_2^2(x_2-1)-r_3^2(x_3-1)]\alpha_s(t_g)h_e(\beta_g,\alpha_a,b_1,b_2)\exp[-S_{gh}(t_g)]\nonumber\\&&-
[r_2^2x_2-r_3^2x_3]\alpha_s(t_h)h_e(\beta_h,\alpha_a,b_1,b_2)\exp[-S_{gh}(t_h)]\}.
\end{eqnarray}

For the CKM suppressed $B_c\rightarrow D^{*+}_{(s)}D^{*0}$ decays,
the decay amplitudes for different polarizations are
\begin{eqnarray}
\mathcal {F}^L_{e2}&=&-2\sqrt{\frac{2}{3}}C_Ff_Bf_{3}\pi M_B^4
\int_0^1dx_2\int_0^{\infty}b_1b_2db_1db_2\phi_{2}(x_2)\exp(-\frac{b_1^2\omega_B^2}{2})\times\nonumber\\&&\{
[-(r_2-2)r_b+2r_2x_2-x_2]\alpha_s(t_a)h_e(\alpha_e,\beta_a,b_1,b_2)S_t(x_2)\exp[-S_{ab}(t_a)]
\nonumber\\&&+r^2_2\alpha_s(t_b)h_e(\alpha_e,\beta_b,b_2,b_1)S_t(x_1)\exp[-S_{ab}(t_b)]\},
\end{eqnarray}
\begin{eqnarray}
\mathcal {F}^N_{e2}&=&-2\sqrt{\frac{2}{3}}C_Ff_Bf_{3}r_3\pi M_B^4
\int_0^1dx_2\int_0^{\infty}b_1b_2db_1db_2\phi_{2}(x_2)\exp(-\frac{b_1^2\omega_B^2}{2})\times\nonumber\\&&\{
[2-r_b+r_2(4r_b-x_2-1)]\alpha_s(t_a)h_e(\alpha_e,\beta_a,b_1,b_2)S_t(x_2)\exp[-S_{ab}(t_a)]
\nonumber\\&&-r_2\alpha_s(t_b)h_e(\alpha_e,\beta_b,b_2,b_1)S_t(x_1)\exp[-S_{ab}(t_b)]\},
\end{eqnarray}
\begin{eqnarray}
\mathcal {F}^T_{e2}&=&2\sqrt{\frac{2}{3}}C_Ff_Bf_{3}r_3\pi M_B^4
\int_0^1dx_2\int_0^{\infty}b_1b_2db_1db_2\phi_{2}(x_2)\exp(-\frac{b_1^2\omega_B^2}{2})\times\nonumber\\&&\{
[2-r_b-r_2(1-x_2)]\alpha_s(t_a)h_e(\alpha_e,\beta_a,b_1,b_2)S_t(x_2)\exp[-S_{ab}(t_a)]
\nonumber\\&&-r_2\alpha_s(t_b)h_e(\alpha_e,\beta_b,b_2,b_1)S_t(x_1)\exp[-S_{ab}(t_b)]\},
\end{eqnarray}
\begin{eqnarray}
\mathcal {M}^L_{e2}&=&\frac{8}{3}C_Ff_B\pi M_B^4
\int_0^1dx_2dx_3\int_0^{\infty}b_2b_3db_2db_3\phi_{2}(x_2)\phi_{3}(x_3)\exp(-\frac{b_1^2\omega_B^2}{2})\times\nonumber\\&&
\{[2-x_1-x_2-x_3-r_2(1-x_2)]\alpha_s(t_c)h_e(\beta_c,\alpha_e,b_3,b_2)\exp[-S_{cd}(t_c)]-
\nonumber\\&&[x_3-x_1+r_2(1-x_2)]\alpha_s(t_d)h_e(\beta_d,\alpha_e,b_3,b_2)\exp[-S_{cd}(t_d)]\},
\end{eqnarray}
\begin{eqnarray}
\mathcal {M}^N_{e2}&=&-\mathcal {M}^T_{e2}=\frac{8}{3}C_Ff_B\pi M_B^4
\int_0^1dx_2dx_3\int_0^{\infty}b_2b_3db_2db_3\phi_{2}(x_2)\phi_{3}(x_3)\exp(-\frac{b_1^2\omega_B^2}{2})\nonumber\\&&\times
\{[r_3(x_1-x_3)]\alpha_s(t_c)h_e(\beta_c,\alpha_e,b_3,b_2)\exp[-S_{cd}(t_c)]+
\nonumber\\&&[2r_c-r_3(1-x_1-x_3)]\alpha_s(t_d)h_e(\beta_d,\alpha_e,b_3,b_2)\exp[-S_{cd}(t_d)]\}.
\end{eqnarray}

\section{Scales and related functions in hard kernel }\label{sec:b}

We show here the functions $h_e$, coming from the Fourier transform
of hard kernel,
\begin{eqnarray}
h_e(\alpha,\beta,b_1,b_2)&=&h_1(\alpha,b_1)\times h_2(\beta,b_1,b_2),\nonumber\\
h_1(\alpha,b_1)&=&\left\{\begin{array}{ll}
K_0(\sqrt{\alpha}b_1), & \quad  \quad \alpha >0\\
K_0(i\sqrt{-\alpha}b_1),& \quad  \quad \alpha<0
\end{array} \right.\nonumber\\
h_2(\beta,b_1,b_2)&=&\left\{\begin{array}{ll}
\theta(b_1-b_2)I_0(\sqrt{\beta}b_2)K_0(\sqrt{\beta}b_1)+(b_1\leftrightarrow b_2), & \quad   \beta >0\\
\theta(b_1-b_2)J_0(\sqrt{-\beta}b_2)K_0(i\sqrt{-\beta}b_1)+(b_1\leftrightarrow b_2),& \quad   \beta<0
\end{array} \right.
\end{eqnarray}
where $J_0$ is the Bessel function and $K_0$, $I_0$ are modified
Bessel function with $K_0(ix)=\frac{\pi}{2}(-N_0(x)+i J_0(x))$. The
hard scale t is chosen as the maximum virtuality of the internal
momentum transition in the hard amplitudes, including
$1/b_i(i=1,2,3)$:
\begin{eqnarray}
t_a&=&\max(\sqrt{|\alpha_e|},\sqrt{|\beta_a|},1/b_1,1/b_2),\quad t_b=\max(\sqrt{|\alpha_e|},\sqrt{|\beta_b|},1/b_1,1/b_2),\nonumber\\
t_c&=&\max(\sqrt{|\alpha_e|},\sqrt{|\beta_c|},1/b_2,1/b_3),\quad t_d=\max(\sqrt{|\alpha_e|},\sqrt{|\beta_d|},1/b_2,1/b_3),\nonumber\\
t_e&=&\max(\sqrt{|\alpha_a|},\sqrt{|\beta_e|},1/b_2,1/b_3),\quad t_f=\max(\sqrt{|\alpha_a|},\sqrt{|\beta_f|},1/b_2,1/b_3),\nonumber\\
t_g&=&\max(\sqrt{|\alpha_a|},\sqrt{|\beta_g|},1/b_1,1/b_2),\quad t_h=\max(\sqrt{|\alpha_a|},\sqrt{|\beta_h|},1/b_1,1/b_2),
\end{eqnarray}
where
\begin{eqnarray}\label{eq:betai}
\alpha_e&=&(1-x_2)(x_1-r_2^2)(1-r_3^2)M_B^2,\quad \alpha_a=-(1+(r_3^2-1)x_2)(1+(r_2^2-1)x_3)M_B^2,\nonumber\\
\beta_a&=&[r_b^2+(r_2^2-1)(x_2+r_3^2(1-x_2))]M_B^2,\quad \beta_b=(1-r_3^2)(x_1-r_2^2)M_B^2,\nonumber\\
\beta_c&=&[r_c^2-(1-x_2(1-r_3^2))(1-x_1-x_3(1-r_2^2))]]M_B^2,\nonumber\\\quad \beta_d&=&(1-x_2)(1-r_3^2)[x_1-x_3-r_2^2(1-x_3)]M_B^2,\nonumber\\
\beta_e&=&-[1+(r_3^2-1)x_2]M_B^2,\quad \beta_f=-[1+(r_2^2-1)x_3]M_B^2,\nonumber\\
\beta_g&=&[r_c^2+(1-x_2(1-r_3^2))(x_1+x_3-1-r_2^2x_3)]M_B^2,\quad \nonumber\\\beta_h&=&[r_b^2-x_2(r_3^2-1)(x_1-x_3(1-r_2^2))]M_B^2.
\end{eqnarray}
The Sudakov factors used in the text are defined by
\begin{eqnarray}
S_{ab}(t)&=&s(\frac{M_B}{\sqrt{2}}x_1,b_1)+s(\frac{M_B}{\sqrt{2}}x_2,b_2)
+\frac{5}{3}\int_{1/b_1}^t\frac{d\mu}{\mu}\gamma_q(\mu)+2\int_{1/b_2}^t\frac{d\mu}{\mu}\gamma_q(\mu),\nonumber\\
S_{cd}(t)&=&s(\frac{M_B}{\sqrt{2}}x_1,b_2)+s(\frac{M_B}{\sqrt{2}}x_2,b_2)+s(\frac{M_B}{\sqrt{2}}x_3,b_3)
\nonumber\\&&+\frac{11}{3}\int_{1/b_2}^t\frac{d\mu}{\mu}\gamma_q(\mu)+2\int_{1/b_3}^t\frac{d\mu}{\mu}\gamma_q(\mu),\nonumber\\
S_{ef}(t)&=&s(\frac{M_B}{\sqrt{2}}x_2,b_2)+s(\frac{M_B}{\sqrt{2}}x_3,b_3)
+2\int_{1/b_2}^t\frac{d\mu}{\mu}\gamma_q(\mu)+2\int_{1/b_3}^t\frac{d\mu}{\mu}\gamma_q(\mu),\nonumber\\
S_{gh}(t)&=&s(\frac{M_B}{\sqrt{2}}x_1,b_1)+s(\frac{M_B}{\sqrt{2}}x_2,b_2)+
s(\frac{M_B}{\sqrt{2}}x_3,b_2),\nonumber\\&&
+\frac{5}{3}\int_{1/b_1}^t\frac{d\mu}{\mu}\gamma_q(\mu)+4\int_{1/b_2}^t\frac{d\mu}{\mu}\gamma_q(\mu),
\end{eqnarray}
where the functions $s(Q,b)$ are defined in Appendix A of \cite{epjc45711}. $\gamma_q=-\alpha_s/\pi$ is the anomalous dimension of the quark.

\section{Meson Wave functions }\label{sec:c}

In the nonrelativistic limit, the $B_c$ meson wave function can be
written as \cite{prd81014022}
\begin{eqnarray}
\Phi_{B_c}(x)=\frac{if_B}{4N_c}[(\rlap{/}{P}+M_{B_c})\gamma_5\delta(x-r_c)]\exp(-\frac{b^2\omega_B^2}{2}),
\end{eqnarray}
in which the last exponent term  represents the $k_T$ distribution.
 Here, we only consider  the dominant Lorentz structure and
neglect another contribution in our calculation \cite{epjc28515}.

In the heavy quark limit, the two-particle light-cone distribution
amplitudes of $D_{(s)}/D_{(s)}^*$ meson are defined as \cite{prd67054028}
\begin{eqnarray}\label{eq:dwave}
\langle D_{(s)}(P_2)|q_{\alpha}(z)\bar{c}_{\beta}(0)|0\rangle &=&
\frac{i}{\sqrt{2N_c}}\int^1_0dx e^{ixP_2\cdot
z}[\gamma_5(\rlap{/}{P}_2+m_{D_{(s)}})\phi_{D_{(s)}}(x,b)]_{\alpha\beta},
\nonumber\\
\langle D_{(s)}^*(P_2)|q_{\alpha}(z)\bar{c}_{\beta}(0)|0\rangle
&=&-\frac{1}{\sqrt{2N_c}}\int^1_0dx e^{ixP_2\cdot
z}[\rlap{/}{\epsilon}_L(\rlap{/}{P}_2+m_{D_{(s)}^*})\phi^L_{D_{(s)}^*}(x,b)\nonumber\\&&
+\rlap{/}{\epsilon}_T(\rlap{/}{P}_2+m_{D_{(s)}^*})\phi^T_{D_{(s)}^*}(x,b)]_{\alpha\beta},
\end{eqnarray}
with the  normalization conditions:
\begin{eqnarray}
\int^1_0dx \phi_{D_{(s)}}(x,0)=\frac{f_{D_{(s)}}}{2\sqrt{2N_c}},\quad  \int^1_0dx
\phi^{L}_{D_{(s)}^*}(x,0)= \int^1_0dx
\phi^{T}_{D_{(s)}^*}(x,0)=\frac{f_{D_{(s)}^*}}{2\sqrt{2N_c}},
\end{eqnarray}
where we have assumed $f_{D_{(s)}^*}=f^T_{D_{(s)}^*}$. Note that
equations of motion do not relate $\phi^L_{D_{(s)}^*}$ and
$\phi^T_{D_{(s)}^*}$. We use the following relations derived from
HQET \cite{hqet} to determine  $f_{D^*_{(s)}}$
\begin{eqnarray}\label{eq:Ddecayc}
f_{D^*_{(s)}}=\sqrt{\frac{m_{D_{(s)}}}{m_{D_{(s)}^*}}}f_{D_{(s)}}.
\end{eqnarray}
The distribution amplitude  $\phi^{(L,T)}_{D_{(s)}^{(*)}}$ is taken as \cite{09101424}
\begin{eqnarray}
\phi^{(L,T)}_{D_{(s)}^{(*)}}=\frac{3}{\sqrt{2N_c}}f_{D^{(*)}_{(s)}}x(1-x)[1+a_{D^{(*)}_{(s)}}(1-2x)]\exp(-\frac{b^2\omega^2_{D_{(s)}}}{2}).
\end{eqnarray}
We use $a_D=0.5 \pm 0.1, \omega_{D}= 0.1 \text{GeV}$ for $D/D^*$ meson and $a_D=0.4 \pm 0.1, \omega_{D_s}= 0.2 \text{GeV}$ for $D_s/D_s^*$ meson,
  which are determined in Ref. \cite{dd1} by fitting.


\end{appendix}


\begin{thebibliography}{99}

\bibitem{iiba}
N. Brambilla et al., (Quarkonium Working Group), CERN-2005-005, hep-ph/0412158.
\bibitem{plb286160}
M. Masetti, Phys. Lett. B \textbf{286}, 160 (1992).
\bibitem{prd62057503}
R. Fleischer and D. Wyler, Phys. Rev. D \textbf{62}, 057503 (2000);
 R. Fleischer, Lect. Notes Phys. \textbf{647}, 42 (2004).
\bibitem{jpg301445}
 V.V. Kiselev, J. Phys. G \textbf{30}, 1445 (2004);
 V.V. Kiselev, A. E. Kovalsky, and A.K. Likhoded, Nucl.
Phys. B \textbf{585}, 353 (2000); V.V. Kiselev, arXiv:hep-ph/
0211021.
\bibitem{plb555189}
 M.A. Ivanov, J.G. K\"{o}rner and O.N. Pakhomova, Phys. Lett.
B \textbf{555}, 189 (2003).
\bibitem{prd65034016}
A. K. Giri, R. Mohanta and M. P. Khanna, Phys. Rev. D \textbf{65}, 034016 (2001).
\bibitem{prd73054024}
 M.A. Ivanov, J.G. K\"{o}rner and P. Santorelli, Phys. Rev. D \textbf{73}, 054024 (2006).
 \bibitem{prd564133}
Jia-Fu Liu  and Kuang-Ta Chao,
Phys. Rev. D \textbf{56}, 4133 (1997).
\bibitem{pan67}
I. P. Gouz, V. V. Kiselev, A. K. Likhoded, V. I. Ro-
manovsky, and O. P. Yushchenko, Phys. Atom. Nucl. \textbf{67},
1559 (2004); Yad. Fiz. \textbf{67}, 1581 (2004).
\bibitem{prd61034012}
P. Colangelo and F. De Fazio, Phys. Rev. D \textbf{61}, 034012
(2000).
\bibitem{prd62014019}
A. Abd El-Hady, J.H. Munoz, and J. P. Vary, Phys. Rev. D
\textbf{62}, 014019 (2000).
\bibitem{prd493399}
C.H. Chang and Y.Q. Chen, Phys. Rev. D \textbf{49}, 3399 (1994).

\bibitem{cheng}
 Hai-Yang Cheng, Chun-Khiang Chua, Phys. Rev. D \textbf{80}, 114008
 (2009).

\bibitem{prl744388}
H.-n. Li, and H.L.Yu, Phys. Rev. Lett. \textbf{74}, 4388 (1995);
H.-n. Li, Phys. Lett. B \textbf{348}, 597 (1995).

\bibitem{plb5046}
Y. Y. Keum, H. n. Li and A. I. Sanda, Phys. Lett. B \textbf{504}, 6 (2001).
\bibitem{prd63074009}
Cai-Dian L\"{u}, K. Ukai and M.-Z. Yang, Phys. Rev. D \textbf{63}, 074009 (2001); Cai-Dian L\"{u} and M.Z.
Yang, Eur. Phys. J. C \textbf{23}, 275 (2002).

\bibitem{scet1}
C.W.Bauer, S. Fleming, D. Pirjol and I. W. Stewart, Phys. Rev. D
\textbf{63}, 114020 (2001); C.W.Bauer, D. Pirjol and I. W. Stewart,
Phys. Rev. Lett. \textbf{87}, 201806 (2001);
 Phys. Rev. D \textbf{65}, 054022 (2002).

\bibitem{09101424}
 Run-Hui Li, Cai-Dian L\"{u}, and Hao Zou,  Phys. Rev. D \textbf{78}, 014018 (2008);
 Hao Zou, Run-Hui Li, Xiao-Xia Wang, and Cai-Dian  L\"{u},  J. Phys. G: Nucl. Part. Phys. \textbf{37}, 015002 (2010).
\bibitem{0512347}
Ying Li, Cai-Dian L\"{u} and Cong-Feng Qiao, Phys. Rev. D
\textbf{73}, 094006 (2006); Ying Li and Cai-Dian L\"{u}, J. Phys. G
\textbf{29}, 2115 (2003);

\bibitem{dd1}
Run-Hui Li, Cai-Dian L\"{u}, A.I. Sanda and Xiao-Xia Wang,  Phys. Rev. D \textbf{81}, 034006 (2010).
\bibitem{dd2}
Ying Li, Cai-Dian L\"{u} and Zhen-Jun Xiao, J. Phys. G \textbf{31},  273 (2005).
\bibitem{npb193381}
J. C. Collins and D. E. Soper, Nucl. Phys. B \textbf{193},
381 (1981); J. Botts and G. Sterman, Nucl. Phys. B  \textbf{325}, 62
(1989).
\bibitem{prd074004}
B. Melic, B. Nizic, and K. Passek, Phys. Rev. D \textbf{60}, 074004
(1999).

\bibitem{epjc45711}
Jian-Feng Cheng, Dong-Sheng Du, and Cai-Dian L\"{u}, Eur. Phy. J. C.
\textbf{45}, 711 (2006).
\bibitem{dpi} Cai-Dian L\"{u}, Phys. Rev. D \textbf{68},
 097502 (2003); Yong-Yeon Keum, et al, Phys.
Rev. D \textbf{69 }, 094018 (2004).


\bibitem{npp37}
 Particle Data Group, J. Phys. G: Nucl.Part. Phys. \textbf{37},  075021 (2010).
\bibitem{prd7905402}
W.Wang,Y.L.Shen and C.D. L\"{u},  Phys. Rev. D \textbf{79}, 054012
(2009).
\bibitem{jpg35085002}
R. Dhir, N. Sharma, and R. C. Verma, J. Phys. G \textbf{35},
085002 (2008); R.C. Verma and A. Sharma, Phys. Rev.
D \textbf{65}, 114007 (2002); R. Dhir and R.C. Verma, Phys. Rev.
D \textbf{79}, 034004 (2009).
\bibitem{prd391342}
Dong-Sheng Du and Z. Wang, Phys. Rev. D \textbf{39}, 1342 (1989).

\bibitem{fac}
Ahmed Ali, G. Kramer, Cai-Dian L\"{u},   Phys. Rev. D \textbf{58},
094009 (1998).
\bibitem{rhorho}
 Ying Li and Cai-Dian L\"{u},        Phys. Rev. D \textbf{73}, 014024 (2006).
\bibitem{11121257}
Zhou Rui, Zhi-Tian Zou and Cai-Dian L\"{u}, arXiv:  1112.1257 [hep-ph].






\bibitem{prd81014022}
Xin Liu, Zhen-Jun Xiao, and Cai-Dian L\"{u}, Phys. Rev. D
\textbf{81}, 014022 (2010).

\bibitem{epjc28515}
  Cai-Dian L\"{u}, M.-Z. Yang, Eur. Phys. J. C \textbf{28}, 515 (2003).
\bibitem{prd67054028}
 T. Kurimoto, H. n. Li and A. I. Sanda,
Phys. Rev. D \textbf{67}, 054028 (2003).
\bibitem{hqet}
 A. V. Manohar and M. B. Wise, Camb. Monogr. Part. Phys. Nucl. Phys. Cosmol. \textbf{10}, 1 (2000).



\end{thebibliography}
\end{document}